\documentclass[preprint,showpacs,preprintnumbers,amsmath,amssymb,double-spaced]{revtex4}
\usepackage{graphicx}
\usepackage{dcolumn}
\usepackage{bm}

\begin{document}
\title{Swap Action in a Solid-State Controllable Anisotropic Heisenberg Model}

\author{Xiang Hao}
\author{Shiqun Zhu}
\altaffiliation{Corresponding author, E-mail: szhu@suda.edu.cn}

\affiliation{School of Physical Science and Technology, Suzhou
University, Suzhou, Jiangsu 215006, People's Republic of China}

\begin{abstract}
Correct swap action can be realized via the control of the
anisotropic Heisenberg interaction in solid-state quantum systems.
The conditions of performing a swap are derived by the dynamics of
arbitrary bipartite pure state. It is found that swap errors can be
eliminated in the presence of symmetric anisotropy. In realistic
quantum computers with unavoidable fluctuations, the gate fidelity
of swap action is estimated. The scheme of quantum computation via
the anisotropic Heisenberg interaction is implemented in a one
dimensional quantum dots. The slanting and static magnetic field can
be used to adjust the anisotropy.
\end{abstract}


\maketitle

\section{Introduction}

The physical implementation of quantum computation is thought of as
a fundamental step for quantum information processing \cite{Ekert96,
Steane98,Bennett00}. In recent years, some proposals have been
offered about the realization of quantum computation, using atoms or
photons in cavity \cite{Turchette95,Sleator95}, trapped ions
\cite{Cirac95,Monroe95}, and bulk NMR techniques
\cite{Cory97,Gershenfeld97}. In the implementation of these quantum
computation architectures, it seems very difficult to perform the
actual large-scale quantum computation. However, the quantum
solid-state computation now attracts a considerable interest because
of the feasible manipulation of many qubits represented by
superconducting Cooper pairs \cite{Shnirnan97}, electron spin in
quantum dots \cite{DiVincenzo98,Burkard99,Imamoglu99}, orbital
energy levels in nanostructures \cite{Bonadeo98}, donor nuclear
spins \cite{Kane98,Privman98} and newly defined pseudospins
\cite{Tokura06}. In principle, the amazing large-scale quantum
computation can be realized in such quantum solid-state systems. The
schemes based on quantum dots have unique advantages in the actual
physical implementation. In quantum dots, the microscopic systems of
two discrete levels can serve as a qubit carrying the elementary
quantum information. The electron spin as a natural two-level
quantum system can be considered as one good qubit with the long
coherence time \cite{Petta05}. The pseudospin using the orbit
degrees of freedom is another one which can be controlled easily
\cite{van03}. More recently, the proposal by combining the spin and
orbit degrees of freedom has been introduced \cite{Tokura06}. Based
on it, coherently controlling single electron spin is possible. It
seems that both good qualities about the long coherence time and
easy manipulation are shown by this newly defined quantum system.
Regardless of the definition of qubit, most of effective interaction
between two coupled qubits are modeled by the Heisenberg exchange
interaction. It has been shown that the universal quantum gates
\cite{Nielsen00} can be realized via the isotropic Heisenberg
interaction and the uniform magnetic field \cite{Burkard99}. In the
realistic spin-based quantum computation, errors from inhomogeneous
Zeeman field \cite{Das01} and anisotropic interactions
\cite{Loss02,Kavokin01} are regarded as a major obstacle in quantum
dots. The errors from inhomogeneous field cannot be eliminated
completely \cite{Das01,Hu03}. Meanwhile, the anisotropy induced by
the spin-orbit coupling can lead to the nontrivial error with the
order of $10^{-4}$ \cite{Kavokin01}. Therefore, it is necessary to
study the method of implementing the quantum computation in the
anisotropic Heisenberg model. The construction of correct swap
action is one crucial step for possible quantum computation.

In this paper, the correct swap action can be performed in the
anisotropic Heisenberg $XXZ$ model. In section II, the conditions of
performing correct swap action are analyzed in detail. The impacts
of the certain fluctuations on swap action are estimated by the gate
fidelity from the Heisenberg interaction, the anisotropy and the
effective Zeeman field. In section III, the possible physical
implementation of such swap action is presented. A discussion
concludes the paper.

\section{Swap Action in Anisotropic Heisenberg Model}

In mangy protocols about the solid-state quantum computation, the
isotropic Heisenberg model is always uesd because the universal
quantum gates can be constructed in this ideal model. Nevertheless,
in realistic quantum computers, there are always anisotropic
exchange interactions. Therefore, the general case of anisotropic
Heisenberg $XXZ$ model needs to be studied. The Hamiltonian of two
coupled qubits $i$ and $j$ can be given as
\begin{equation} \label{eq1}
H_{ij}=J(S_{i}^{x}S_{j}^{x}+S_{i}^{y}S_{j}^{y}+\Delta
S_{i}^{z}S_{j}^{z})+\Gamma(S_{i}^{z}+S_{j}^{z})
\end{equation}
where $S_{i}^{\alpha}=\frac 12\sigma_{i}^{\alpha}(\alpha=x,y,z)$ are
three components of qubit $i$ operator and $\sigma_{i}^{\alpha}$ is
the Pauli operator. For the convenience of computation,
$|0\rangle_{i}, |1\rangle_{i}$ are assumed to be the eigenstates of
$\sigma_{i}^{z}$ with the corresponding eigenvalues $\pm 1$, $J$ is
the effective strength of Heisenberg exchange interaction and
$\Delta$ is the anisotropy parameter. If the external magnetic field
$\vec{B}$ is along $z$ direction, $\Gamma$ is the effective Zeeman
splitting energy with $\Gamma=g\mu_{B}B$, $g$ is the effective $g$
factor and $\mu_{B}$ is the Bohr magneton. For the possible
realization of quantum computation, the parameters $J, \Delta$, and
$\Gamma$ can be tunable in quantum computers.

To show the dynamics of the anisotropic Heisenberg $XXZ$ model, the
eigenstates $|\psi\rangle$ and corresponding eigenvalues $E$ of
$H_{ij}$ need to be derived. In this model, the total spin is
conserved since $[H_{ij},S_{i}^{z}+S_{j}^{z}]=0$. In the product
space of two qubits, $|\psi\rangle$ and $E$ can be easily obtained
by $E_1=\frac {J\Delta}4+\Gamma, E_2=\frac {J\Delta}4-\Gamma,
E_3=-\frac {J(\Delta-2)}4, E_4=-\frac {J(\Delta+2)}4$ and
$|\psi_1\rangle=|00\rangle, |\psi_2\rangle=|11\rangle,
|\psi_3\rangle=\frac 1{\sqrt 2}(|01\rangle+|10\rangle),
|\psi_4\rangle=\frac 1{\sqrt 2}(|01\rangle-|10\rangle)$. The general
unitary transformation on qubits $i$ and $j$ can be expressed by
\begin{equation}\label{eq2}
U_{ij}(t)=T\exp\{-i\int_{0}^{t}H_{ij}(t')dt'\}
\end{equation}
where $T$ is the time ordering operator. The swap action is just one
of the unitary operations $U_{sw}$ by which the states at qubit $i$
and $j$ can be exchanged. Without losing of the generality, an
initial arbitrary quantum product state of qubits $i$ and $j$ can be
assumed to be
\begin{equation}\label{eq3}
|\Psi_{in}\rangle=(\alpha_1|0\rangle_i+\alpha_2|1\rangle_i)\otimes(\beta_1|0\rangle_j+\beta_2|1\rangle_j)
\end{equation}
where $\alpha_1,\alpha_2,\beta_1,\beta_2$ are arbitrary complex
coefficients which satisfy $|\alpha_1|^2+|\alpha_2|^2=1$ and
$|\beta_1|^2+|\beta_2|^2=1$. In the process of time evolution, the
reduced density matrix $\rho_i$ and $\rho_j$ are generally turned
into the mixed ones, i. e., $\rho_{ij}\neq \rho_i\otimes\rho_j$,
which is useless for the setup of the swap action. Only if the
density matrix $\rho_{ij}$ can be expressed by the product of
$\rho_i$ and $\rho_j$, it can be used to construct the swap action.
The method is very crucial by which the pure state $\rho_i$ and
$\rho_j$ can be determined at a certain time. To obtain the swap
action, the theorem about an arbitrary $2\times2$ matrix $A$ is
introduced by
\begin{equation}\label{eq4}
A^2-Tr(A)A=Det(A)I
\end{equation}
where $Tr(A)$ is the trace norm of matrix $A$, $Det(A)$ is the
determinant of $A$ and $I$ is the unity matrix. If $A$ is a pure
state, the determinant satisfies $Det(A)\equiv0$. Thus, by means of
calculating the determinant of $\rho_i$ or $\rho_j$, the reduced
density matrix possibly denotes a pure state at certain time when
the determinant is zero. A swap action can be finally constructed.
The quantum state $|\Psi(t)\rangle$ at time $t$ is given by
\begin{align}\label{eq5}
|\Psi(t)\rangle&=\alpha_1\beta_1e^{-i(\phi_H+\phi_Z/4)}|00\rangle+\frac
12[\gamma_1e^{i(\phi_Z/4-\phi_X/2)}+\gamma_2e^{i(\phi_Z/4+\phi_X/2)}]|01\rangle&\nonumber
\\
&+\alpha_2\beta_2e^{i(\phi_H-\phi_Z/4)}|11\rangle+\frac
12[\gamma_1e^{i(\phi_Z/4-\phi_X/2)}-\gamma_2e^{i(\phi_Z/4+\phi_X/2)}]|10\rangle&
\end{align}
where $\gamma_1=\alpha_1\beta_2+\alpha_2\beta_1$ and
$\gamma_2=\alpha_1\beta_2-\alpha_2\beta_1$. The phase angles are
given by $\phi_H=\int_{0}^{t}\Gamma dt'$,
$\phi_Z=\int_{0}^{t}J\Delta dt'$ and $\phi_X=\int_{0}^{t}Jdt'$. The
reduced density matrix $\rho_i$ can be easily obtained by
\begin{equation}\label{eq6}
\rho_i(t)=\left(\begin{array}{cc}
a_{00} & a_{01} \\
a_{10} & a_{11}
\end{array}\right)
\end{equation}
Here the elements of the matrix are calculated as
\begin{align}\label{eq7}
a_{00}&=|\alpha_1\beta_1|^2+\frac
14(|\gamma_1|^2+|\gamma_2|^2+\gamma_1\gamma_2^{\ast}e^{-i\phi_X}+\gamma_1^{\ast}\gamma_2e^{i\phi_X})
\nonumber \\
a_{01}=a_{10}^{\ast}&=\frac
12\alpha_1\beta_1e^{-i(\phi_Z/2+\phi_H)}(\gamma_1^{\ast}e^{i\phi_X/2}-\gamma_2^{\ast}e^{-i\phi_X/2})
\nonumber \\
&+\frac
12\alpha_1\beta_1e^{i(\phi_Z/2-\phi_H)}(\gamma_1e^{-i\phi_X/2}-\gamma_2e^{i\phi_X/2})
\nonumber \\
a_{11}&=|\alpha_2\beta_2|^2+\frac
14(|\gamma_1|^2+|\gamma_2|^2-\gamma_1\gamma_2^{\ast}e^{-i\phi_X}-\gamma_1^{\ast}\gamma_2e^{i\phi_X})
\end{align}
The determinant is given by $Det(\rho_i)=a_{00}a_{11}-a_{01}a_{10}$
and can be simplified by
\begin{equation}\label{eq8}
Det(\rho_i)=\left|\alpha_1\alpha_2\beta_1\beta_2-\frac
14[\gamma_1^2e^{i(\phi_Z-\phi_X)}-\gamma_2^2e^{i(\phi_Z+\phi_X)}]\right|^2
\end{equation}
If the density matrix $\rho_i$ is a pure state, the determinant is
zero. For arbitrary complex coefficients
$\alpha_1,\alpha_2,\beta_1,\beta_2$, the condition of
$Det(\rho_i)\equiv0$ needs the phase angles which simultaneously
satisfy $\phi_Z-\phi_X=2n\pi $ and $\phi_Z+\phi_X=2m\pi $ with
$m\neq n$ and $m,n=0,\pm 1, \pm 2,\cdots$. That is,
$\phi_Z=(m+n)\pi$ and $\phi_X=(m-n)\pi$. Furthermore, when the value
of $|m-n|$ is even at certain time $\tau_{o}$, the state of qubit
$i$ can be expressed by
$|\psi_i(\tau_{o})\rangle=\alpha_1|0\rangle_i+\alpha_2\exp[i(\pi
n+\phi_H(\tau_{o}))]|1\rangle_i$ with the additional phase to the
original state $|\psi_i(0)\rangle$. When the value of $|m-n|$ is odd
at another certain time $\tau_{s}$, the state is
$|\psi_i(\tau_{s})\rangle=\beta_1|0\rangle_i+\beta_2\exp[i(\pi
n+\phi_H(\tau_{s}))]|1\rangle_i$. If the phase angle
$\phi_H(\tau_{s})=n\pi$,
$|\psi_i(\tau_{s})\rangle=|\psi_j(0)\rangle$, which is just the
original state of qubit $j$. The correct swap action can be
performed at this moment. Similarly,
$|\psi_i(\tau_{o})\rangle=|\psi_i(0)\rangle$ at
$\phi_H(\tau_{o})=n\pi$ and the state of qubit $i$ keeps invariant.
In some solid-state quantum computation architecture, the anisotropy
$\Delta$ can be adjusted by the time independent parameter. Thus, a
correct swap action $U_{sw}$ will be set up if the conditions are
satisfied by
\begin{equation}\label{eq9}
\int_{0}^{\tau_{s}}J\Delta dt'=(m+n)\pi, \quad \int_{0}^{\tau_{s}}J
dt'=(m-n)\pi, \quad \int_{0}^{\tau_{s}}\Gamma dt'=n\pi
\end{equation}
By combining the corresponding single qubit operations and the swap
action, the other two-qubit gate like CNOT gate will be easily
constructed \cite{DiVincenzo98}.

In the previous work \cite{Yin05}, Yin et. al. discussed the
Heisenberg $XXZ$ model for quantum swap action. By means of the time
evolution of one single-qubit reduced density matrix, they found
that the Heisenberg XXZ model of $0\leq\Delta<1$ cannot be used to
perform the exact swap action. However, the anisotropy $\Delta>1$
needs to be included in the general Heisenberg $XXZ$ model.
Different from the result of \cite{Yin05}, the general solution of a
swap action is obtained by Eq. (4) when the anisotropy is
$\Delta>1$. From Eq. (9), it is clear that the anisotropy of
$0\leq\Delta<1$ cannot be used to perform a swap action. This is
consistent with the previous work \cite{Yin05}.

Apart from the case of Eq. (9), the swap errors cannot be neglected.
In realistic quantum computation, certain fluctuations
\cite{Vidal99,Bandyopadhyay04} from internal and external impacts
are unavoidable. For quantum gates, the fluctuations from
$\phi_X,\phi_Z$ and $\phi_H$ will influence the performance of
$U_{sw}$. For the weak coupling, the Gaussian distributions of
$\phi_X\sim N(\overline{\phi}_X,\lambda_X), \phi_Z\sim
N(\overline{\phi}_Z,\lambda_Z)$ and $\phi_H\sim
N(\overline{\phi}_H,\lambda_H)$ can be reasonably assumed. The
Gaussian distribution $N(\overline{\phi},\lambda)=\frac
1{\sqrt{2\pi}\lambda}e^{-(\phi-\overline{\phi})^2/2\lambda^2}$ with
the mean value $\overline{\phi}$ and the standard deviation
$\lambda$. These fluctuations are possible attributable to those of
exchange interaction $J$, anisotropy $\Delta$ and the effective
Zeeman field $\Gamma$. To evaluate the effects of such fluctuations
on $U_{sw}$, the gate fidelity \cite{Poyatos97} can be introduced as
$F=\overline{\langle \Psi_{in}|U_{sw}^{\dag}\rho
U_{sw}|\Psi_{in}\rangle}$ where the overline denotes the average for
all the possible initial states. After the calculation, the general
fidelity is expressed by
\begin{equation}\label{eq10}
F(\phi_X,\phi_Z,\phi_H)=\frac 15+\frac {8}{15}\sin^2\frac
{\phi_X}2+\frac {4}{15}\sin\frac {\phi_X}2\sin(\frac
{\phi_Z}2+\phi_H)
\end{equation}
When the mean value of the distributions is chosen to be those given
by Eq. (9), the average fidelity $F_A$ with fluctuations is obtained
by
\begin{equation}\label{eq11}
F_A=\frac {7}{15}+\frac
{4}{15}(e^{-\lambda^2_X/2}+e^{-(\lambda^2_X+\lambda^2_Z+4\lambda^2_H)/8})
\end{equation}
It is seen that the fluctuation from $\phi_X$ mainly determined by
the Heisenberg interaction $J$ is always dominate in contrast to
others. If the deviations $\lambda_X,\lambda_Z,\lambda_H\rightarrow
\infty$, the limit of the average fidelity is $F_A \rightarrow \frac
{7}{15}$. In Fig. 1, the condition of $\lambda_X=\lambda_Z$ is
clearly shown. The gate fidelity is decreased more rapidly with
$\lambda_X$ than that with $\lambda_H$.

\section{Implementation Based on One Pseudospin}

From one kind of newly defined pseudospin \cite{Tokura06}, the
implementation of quantum computation can be possibly performed via
the controllable anisotropic Heisenberg model in quantum dots. As an
effective qubit, this pseudospin can be constructed in a
$z$-directional quantum dot. The either end of the quantum dot is
applied by a ferromagnetic gate electrodes that creates a magnetic
field gradient $\vec{b}$ along $x$ axis. If another external
magnetic field $\vec{B}_0$ is applied along $z$ axis, the total
slanting magnetic field is $\vec{B}=B_0\vec{e}_z+zb\vec{e}_x$. It is
noticed that the slanting magnetic field is static and tunable. The
Hamiltonian of single electron in the parabolic confinement
potential like $GaAs$ can be expressed by
\begin{equation}\label{eq12}
h_i=-\frac {\hbar^2}{2m}\frac {d^2}{dz^2}+\frac
{m\omega^2_0z^2}2+g\mu_B(B_0S^{z}_{i}+zbS^{x}_{i})
\end{equation}
where $m$ is the effective mass, $\omega_0$ is the frequency of the
potential. The last term at the right hand side of Eq. (12) is the
Zeeman splitting energy from the slanting magnetic field $\vec{B}$.
The amplitude of $\vec{B}$ is quite large that the effect of the
Zeeman splitting energy cannot be neglected. If $|g\mu_BB|<\hbar
\omega_0$, the effective two-level quantum system can be formed at
the ground state. In general, the amplitude of field gradient
$\vec{b}$ is smaller than that of external field $\vec{B}_0$. Thus,
it is reasonable to apply the perturbation method to the splitting
of $H'_{i}=g\mu_BzbS^{x}_{i}$. For the convenience, the length of
the confinement potential is chosen as $L=\sqrt{\frac
{2\hbar}{m\omega_0}}$. In the product space of spin and orbit degree
of freedom $\{|n,s\rangle,n=0,1,2,\cdots, s=\pm \}$, the
ground-state energy $E_{0,s}$ can be given by the second order
approximation,
\begin{equation}\label{eq13}
E_{0,s}\approx E^{(0)}_{0,s}+\frac
{\left|\langle0,s|H'_{i}|1,-s\rangle\right|^2}{E^{(0)}_{0,s}-E^{(0)}_{1,-s}}
\end{equation}
where $E^{(0)}_{n, s}=(n+\frac 12)\hbar \omega_0+g\mu_BB_0s$ is the
zeroth order energy and $\langle0,s|H'_{i}|1,-s\rangle=-\frac{\sqrt
2}2g\mu_BbL$. Simultaneously, the ground state is calculated by the
first order approximation,
\begin{equation}\label{eq14}
|\varphi(0,s)\rangle \approx |0,s\rangle+C_{is}|1,-s\rangle
\end{equation}
where the coefficient for pseudospin $i$ is $C_{is}=\frac
{\langle0,s|H'_{i}|1,-s\rangle}{E^{(0)}_{0,s}-E^{(0)}_{1,-s}}$. The
two-level states of this pseudospin are described by
$|\varphi(0,+)\rangle,|\varphi(0,-)\rangle$ with the corresponding
splitting energy $E_{0,+},E_{0,-}$. Therefore, the effective
Hamiltonian for this pseudospin $i$ can be written as
$h_{eff}=\omega S^{z}_{i}$ where the transition frequency
$\omega=|E_{0,s}-E_{0,-s}|$.

Based on this pseudospin, two coupled quantum dots $i$ and $j$ can
be constructed in series \cite{van03}. After the introduction of the
tunneling and inter-dot interaction, the effective Hamiltonian
$H_{eff}$ mapped into the qubits is obtained by \cite{Tokura06}
\begin{equation}\label{eq15}
H_{eff}=J_{eff}(S^{x}_{i}S^{x}_{j}+S^{y}_{i}S^{y}_{j}+
\widetilde{\Delta}S^{z}_{i}S^{z}_{j})+\widetilde{\omega}(S^{z}_{i}+S^{z}_{j})
\end{equation}
where the effective interaction $J_{eff}=\frac {4t_{+}t_{-}}{U-V}$,
the effective anisotropy $\widetilde{\Delta}=\frac
{t^2_{+}+t^2_{-}}{2t_{+}t_{-}}-\frac
{f^2}{t_{+}t_{-}[1-\omega^2/(U-V)^2]}$ and the effective Zeeman
splitting $\widetilde{\omega}=\omega[1-\frac
{2f^2}{(U-V)^2-\omega^2}]$ with $f=\frac 12(f_{+}+\frac
{g_jb_j}{g_ib_i}f_{-})$. It is noted that the inhomogeneity from
$g_i\neq g_j$ and $b_i\neq b_j$ is considered. The parameter $U$ is
the charge energy, $V$ is the strength of the inter-dot interaction,
$t_{\pm}$ and $f_{\pm}$ are the tunneling terms. The expressions of
$t_{\pm}$ and $f_{\pm}$ are given by
$t_{\pm}=t_{00}+C_{i\pm}C_{j\pm}t_{11}$ and
$=(C_{i\pm}+C_{j\mp})t_{12}$ where $t_{mn}$ is the tunneling
amplitude from level $m$ in dot $i$ to level $n$ in dot $j$.
Although the parameters of the effective Hamiltonian are complicate,
the Hamiltonian is the anisotropic Heisenberg $XXZ$ model discussed
in Sec. II. The effective interaction $J_{eff}$, the anisotropy
$\widetilde{\Delta}$ and the Zeeman splitting energy
$\widetilde{\omega}$ can be adjusted via the static slanting
magnetic field. When the conditions given in Eq. (9) are satisfied,
the correct swap action can be performed in this quantum computer.
It is also shown that the study of how to perform a swap action in
anisotropic Heisenberg model is very instructive.

\section{Discussion}

The correct swap action can be performed in the anisotropic
Heisenberg $XXZ$ model via the control of anisotropic interactions
and the effective Zeeman field. The conditions of performing a
perfect swap action are derived by the dynamics of arbitrary
bipartite pure initial state. Considering the fluctuations in
realistic quantum computers, the gate fidelity is used to estimate
the robust ability of swap action against noise. It is found that
the impact of the phase fluctuations $\phi_X$ from the Heisenberg
interaction is dominant in contrast to those of $\phi_Z$ and
$\phi_H$. Based on the newly introduced pseudospin \cite{Tokura06},
the possible physical realization of swap action is illustrated.
Swap errors can be eliminated in the model of tunable anisotropic
interactions.

\begin{acknowledgements}
It is a pleasure to thank Yinsheng Ling for the enlightening
discussion of the topic. The financial support from the Specialized
Research Fund for the Doctoral Program of Higher Education of China
(Grant No. 20050285002) is gratefully acknowledged.
\end{acknowledgements}

\newpage

{\Large Fig. 1}

The average gate fidelity is plotted when $\lambda_X=\lambda_Z$.

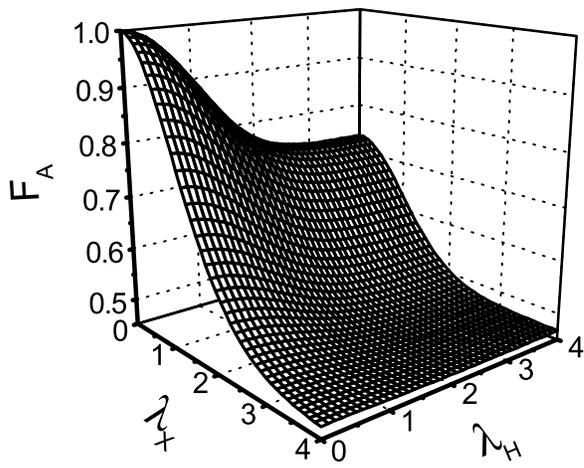
\end{document}